\documentclass[twocolumn,jcp,superscriptaddress]{revtex4}
\usepackage{graphicx}
\usepackage[usenames]{color}
\usepackage{amssymb}

\usepackage{amsmath}
\usepackage{amsfonts}
\usepackage{mathrsfs}

\renewcommand{\epsilon}{\varepsilon}
\newcommand{\figurewidth}{0.45\textwidth}
\newcommand{\narrowfigurewidth}{0.45\textwidth}

\begin{document}

\title{Dynamics of polymer translocation into an anisotropic confinement}

\author{Kehong Zhang}

\affiliation{CAS Key Laboratory of Soft Matter Chemistry, Department
of Polymer Science and Engineering, University of Science and
Technology of China, Hefei, Anhui Province 230026, P. R. China}
\affiliation{College of Light-Textile Engineering and Art, Anhui
Agricultural University, Hefei, Anhui Province 230036, P. R. China}

\author{Kaifu Luo}
\altaffiliation[]{Author to whom the correspondence should be
addressed} \email{kluo@ustc.edu.cn} \affiliation{CAS Key Laboratory
of Soft Matter Chemistry, Department of Polymer Science and
Engineering, University of Science and Technology of China, Hefei,
Anhui Province 230026, P. R. China}

\date{\today}

\begin{abstract}

Using Langevin dynamics simulations, we investigate the dynamics of a flexible polymer translocation
into a confined area under a driving force through a nanopore. We choose an ellipsoidal shape for the
confinement and consider the dependence of the asymmetry of the ellipsoid measured by the aspect ratio
on the translocation time.
Compared with an isotropic confinement (sphere), an anisotropic confinement (ellipsoid) with the same volume slows down
the translocation, and the translocation time increases with increasing the aspect ratio of the ellipsoid.
We further find that it takes different time for polymer translocation into the same ellipsoid
through major-axis and minor-axis directions, depending on the average density of the whole chain in the ellipsoid, $\phi$.
For $\phi$ lower than a critical value $\phi_c$, the translocation through minor axis is faster, and vice versa.
These complicated behaviors are interpreted by the degree of the confinement and anisotropic confinement induced folding
of the translocated chain.

\end{abstract}


\maketitle

\section{INTRODUCTION}

The transport of biopolymers through nanopores is a challenging problem both for polymer physics and molecular biology.
It has received great attention in both experimental and theoretical studies due to its important role in many crucial
biological processes, such as mRNA translocation across nuclear pores, the threading of proteins through biological
membrane channel \cite{Alberts,Kasianowicz1,Kasianowicz2,Meller1,Sung1,Chuang1,Lubensky1,Muthukumar1,Muthukumar2,Tsuchiya1,
Slonkina1,Kasianowicz3,Smith1}, and the viral injection of DNA into host cells \cite{Evilevitch1,Jose1,Frutos1,
Lof1}. Recent experiments show that the translocation processes have potential biotechnological applications, including
rapid DNA or RNA sequencing \cite{Kasianowicz1,Meller1,Meller2,Meller3,Branton1}, gene therapy \cite{Hanss1},
filtration, and controlled drug delivery \cite{Holowka1}.

Due to the loss of the accessible degrees of freedom, the passage of polymer through a nanopore faces a large entropic
barrier and needs an external force to overcome it. The driving force can be provided by an external applied electric
field in the pore \cite{Luo1,Luo2,Luo3,Kantor1,Luo4,Luo5,Dubbeldam1,Vocks1,Dubbeldam2,Gauthier1,Luo6,Cohen1,
Bhattacharya2,Sakaue1,Kasianowicz4}, a pulling force exerted on the end of a polymer \cite{Kantor1,Luo7}, a chemical
potential, binding particles (chaperones) \cite{Metzler1,Metzler2,Zandi1,Luo8,Abdolvahab1}, longitudinal drag flow
in a nanochannel  \cite{Luo12}, or geometrical confinement of the
polymer \cite{Luo11}.

Most of previous studies focused on translocation into an unconfined \emph{trans} side.
The dynamics of polymer translocation into a confined space through a nanopore is of great importance, because it is
related to many biological processes, such as the bacterial gene swapping and
DNA prepackaging \cite{deGennes2}. Experimentally, Smith \emph{et al.} \cite{Smith2}
have found that the rate of packaging is constant until about $50\%$ of the DNA is packaged, then dramatically decreases.
In addition, they have also observed the pause during the packaging process.
Using Langevin dynamics simulations we have investigated the dynamics of polymer translocation
into the regions between two parallel plane walls and a fluidic channel \cite{Luo10}. Compared with an unconfined environment,
the translocation dynamics is greatly changed due to the entropic resisting force induced by crowding effect of the partially
translocated monomers.
For polymer translocation into a closed confinement, experimental results for the rate of packaging
and the pause \cite{Smith2} have been confirmed by stochastic rotation dynamics simulations \cite{Ali1,Ali2,Ali3}
and Langevin dynamics simulations \cite{Muthukumar3}.
Most recently, using Langevin dynamics simulations we have investigated the dynamics of polymer translocation into a circular
nanocontainer through a nanopore under a driving force \cite{Zhang}. We have further found that the translocation time distribution and
the scaling exponent of the translocation time as chain length depend on the average density of the whole chain in the
nanocontainer.
Certainly, for the prepackaging of DNA into the virus capsid, electrostatic charges should be considered.
As noted by Ali \emph{et al.} \cite{Ali4}, when DNA charge is less screened or the Debye length increases,
the packing becomes more difficult and may stop midway.

Interestingly, Ali \emph{et al.} \cite{Ali2} have already observed that capsid geometries play very important role in the packaging
and ejection dynamics of polymers. For a flexible polymer a sphere packs more quickly and ejects more slowly than an ellipsoid.
However, in their simulations the lengths of the major axis and minor axis of the ellipsoid are fixed, and thus it is still not
clear how the aspect ratio (anisotropy) of the ellipsoid affects the packaging dynamics. In addition, in their simulations polymer
translocation into the ellipsoid only through its major axis is explored. An interesting question is what is the difference for
polymer packaging into the same ellipsoid through its major axis and minor axis?

To this end, using Langevin dynamics (LD) simulations, we investigate the dynamics of a flexible polymer translocation through nanopores
into an anisotropic confinement. We emphasize the influence of the aspect ratio of the ellipsoid and translocation direction on the
translocation dynamics.

The paper is organized as follows. In Sec. II, we briefly describe our model and the simulation technique. In Sec.
III, we present our results. Finally, the conclusions and discussion are presented in
Sec. IV.

\section{MODEL AND METHODS}

In our simulations, the polymer is a coarse-grained chain of Lennard-Jones (LJ) particles with the finite extension nonlinear
elastic (FENE) potential \cite{Kremer}. The excluded volume interaction between beads is generated by a short range repulsive
LJ potential:
\begin{equation}
        U_{LJ}=4\varepsilon[(\frac{\sigma}{r})^{12} - (\frac{\sigma}{r})^6] + \varepsilon,
\end{equation}
for $r \leq 2^{1/6}\sigma$ and 0 for $r > 2^{1/6}\sigma$. Here, $\sigma$ is the diameter of a bead and $\varepsilon$ is the
depth of potential. The connectivity between neighboring beads is modeled as a FENE spring with
\begin{equation}
     U_{FENE}(r)=-\frac{1}{2}kR_0^2\ln(1-r^2/R_0^2),
\end{equation}
where $r$ is the distance between consecutive beads, $k$ is the spring constant and $R_0$ is the maximum allowed separation
between connected beads.

\begin{figure}
\includegraphics*[width=\figurewidth]{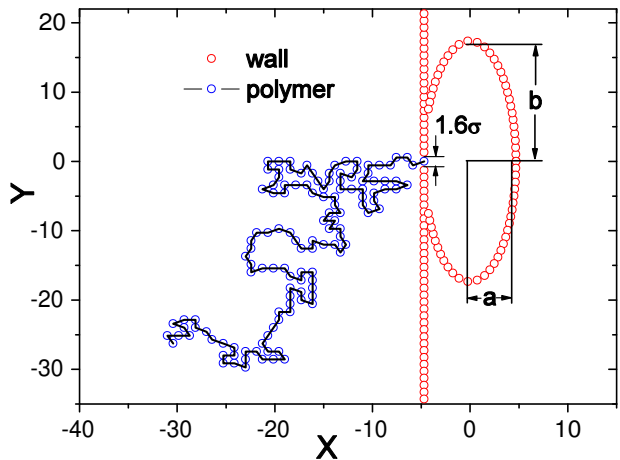}
\caption{Schematic representation of polymer translocation through a nanopore into two-dimensional confined ellipsoid under
an external driving force. The width of the pore is $1.6\sigma$. The lengths of the semi-axis on $x$ and $y$ directions are $a$ and $b$, respectively
        }
\label{Fig1}
\end{figure}

We consider a two-dimensional geometry as depicted by Fig. \ref{Fig1}, where the process of polymer translocation into a confinement through a
nanopore driven by a transmembrane electric field (or molecular motor for real biological system) is shown. The shape of the confinement is described by
\begin{equation}
     x^{2}/a^{2} + y^{2}/b^{2} = 1,
\end{equation}
where $a$ is the length of semi-axis along the $x$ axis and $b$ is the length of semi-axis along the $y$ axis.
The confinement is a two-dimensional sphere for $a=b$ and a two-dimensional ellipsoid for $a\neq b$.

The setup of the confinement is modeled as a hard shell with a pore of width $w =1.6\sigma$, which is formed by stationary wall particles within a
distance $\sigma$ from each other. The entrance permits the translocation of one monomer at a time. Between all monomer-wall
particle pairs, there exits the same short range repulsive LJ interaction as shown above. In Langevin dynamics simulation,
each monomer is subjected to conservative, frictional and random forces, respectively,
\begin{equation}
        m{\bf \ddot{r}}_i=-{\bf\nabla}({U}_{LJ}+{U}_{FENE})+{\bf F}_{\textrm{ext}}-\xi {\bf v}_i+{\bf F}_i^R,
\end{equation}
where $m$ is the monomer's mass, $\xi$ is the friction coefficient, ${\bf v}_i$ is the bead's velocity, and ${\bf F}_i^R$ is the
random force satisfying the fluctuation-dissipation theorem \cite{Allen}.
The external driving force is expressed as ${\bf F}_{ext} = F\widehat{x} $, where $F$ is the external force strength exerted exclusively
on the beads in the pore, and $\widehat{x}$ is a unit vector in the direction along the pore axis.

In the present work, we use the LJ parameters $\varepsilon$, $\sigma$, and bead mass $m$ to fix the system energy, length, and mass scales,
respectively. The time scale is given by $t_{LJ} = (m\sigma^{2}/\varepsilon)^{1/2}$, and the force scale is $\epsilon/\sigma$. To mimic
single-stranded DNA, we choose $\sigma \approx 1.5$ nm \cite{Smith3,Murphy} and the bead mass $m\approx 936$ amu \cite{Luo9}. We set
$k_{B}T = 1.2\varepsilon$, so the interaction strength $\varepsilon$ corresponds to $3.39\times10^{-21}$ J at actual temperature 295 K.
Thus, this leads to a time scale of 32.1 ps and a force scale of 2.3 pN \cite{Luo9}. In our simulations, the dimensionless parameters
are chosen to be $R_0 = 2$, $k = 7$ and $\xi = 0.7$ unless otherwise stated. Corresponding to the range of voltages used in the
experiments \cite{Kasianowicz1, Meller1, Meller2}, the driving force $F$ is set between 2 and 15. The Langevin equation is integrated in
time by the method proposed by Ermak and Buckholz \cite{Ermak}.

Initially, the first monomer is placed at the pore center, while the
remaining monomers are undergoing thermal collisions described by the Langevin thermostat to reach the equilibrium state of the
system. Typically, each simulation data is
the average result of over 1000 successful translocation events to minimize statistical errors. Here, we defined the successful
translocation as the event when the whole chain enters into the confinement.

\section{RESULTS AND DISCUSSIONS}

\subsection{Theory}

Although the chain conformation of a polymer confined in a slit or a tube has been broadly studied \cite{ deGennes3,Cassasa1,deGennes1,Rubinstein1,Cacciuto1,Sakaue2,
Micheletti1}, it is still a challenging problem for a polymer confined to a closed anisotropic system. For a polymer of chain length $N$ confined
to an ellipsoid, we define a parameter $\delta_{ab}$ to characterize the aspect ratio (anisotropy) of the ellipsoid,

\begin{equation}
        \delta_{ab} = b/a.
\end{equation}

The density of beads in the two-dimensional ellipsoid is
\begin{equation}
        \phi = N{\sigma}^{2}/{4ab}.
\end{equation}
According to the blob picture, a confined chain forms a self-avoiding walk consisting of $n_b$ blob with size $\xi_b$.
Each blob contains $g=C{(\xi_b/\sigma)}^{1/\nu_{2D}}$ beads, where $\nu_{2D}=3/4$ being the Flory exponent in two dimensions \cite{deGennes1,Rubinstein1}
and $C$ is a constant, depending on the temperature and the solvent.
Then, the density of beads in a blob is
\begin{equation}
        \phi_{blob}=g\sigma^2/\xi_b^2,
\end{equation}
Based on the relation of $\phi_{blob}=\phi$, we obtain the number of blobs
\begin{equation}
        n_b=N/g=\frac{1}{C}N\phi^{1/(2\nu_{2D}-1)}=\frac{1}{C}N\phi^2.
\end{equation}
Thus, for a polymer with a fixed chain length $N$, $n_b$ is determined only by the area of the confinement ($\pi ab$),
independent of its shape.
Certainly, the above scaling arguments can only work as long as the minor axis is larger than the blob diameter.

For polymer confined in an isotropic space, the free energy cost is proportional to the number of blobs. In units of $k_BT$, it is
\begin{equation}
        \mathcal{F} \sim N\phi^2,
\end{equation}
which indicates that the free energy only depends on the area of the confinement for a fixed chain length.
However, for polymer confined in an ellipsoid with the same area as the sphere, the blob tends to orient along the major axis.
Thus, the free energy $\mathcal{F}$ should include another term for the blob orientation which depends on the shape of the ellipsoid.
Compared with polymer confined in sphere, the free energy for polymer confined in ellipsoid is higher.
For polymer packaging, once the translocated chain is confined, the resisting force should be larger for the ellipsoid than that for
the sphere with the same volume. As a result, an ellipsoid packs more slowly than a sphere, and with increasing the aspect ratio of the
ellipsoid the packaging rate should decrease.
Certainly, for nonequilibrated packaging process the shape dependent free energy implies more complicated translocation dynamics.

During the translocation process,  with increasing the translocated monomers the entropic resisting force $f(\phi(t))$ increases
and depends on the shape of the confinement. Owing to the highly nonequilibrium nature of the translocation process and the dependence
of the shape of the confinement, it is difficult to calculate $f(\phi(t))$. Here, we can write the translocation time as \cite{Zhang}
\begin{equation}
\tau \sim N^\alpha/[F-f(\phi)]=N^\alpha/[F(1-f(\phi)/F)],
\label{eq6}
\end{equation}
where $\alpha$ is the scaling exponent of the translocation time $\tau$ with chain length $N$, and $F(1-f(\phi)/F)$ is the effective
driving force. For an unconfined translocation, translocation time is
\begin{equation}
\tau_\infty \sim N^\alpha/F
\label{eq8}
\end{equation}
for weak driving forces, and then we obtain
\begin{equation}
1-\frac{\tau_\infty}{\tau} \sim f(\phi)/F. \label{eq9}
\end{equation}
Based on this relationship, we can explore the dependence of the average resisting force $f(\phi)$ on the density of the chain $\phi$
and the shape of the ellipsoid.

\subsection{Numerical results}

In our simulations, $\delta_{ab}=1$ corresponds to translocation into a sphere.
The shape of the ellipsoid does not change but its orientation changes for the pairs
of $\delta_{ab}=0.125$ and 8, $\delta_{ab}=0.25$ and 4, $\delta_{ab}=0.5$ and 2, respectively.
Here, $\delta_{ab}>1$ and $\delta_{ab}<1$ denote translocation
into the ellipsoid along the minor-axis and the major-axis directions, respectively.

\subsubsection{Translocation probability as a function of the driving force and the density of the chain}

\begin{figure}
\includegraphics*[width=\figurewidth]{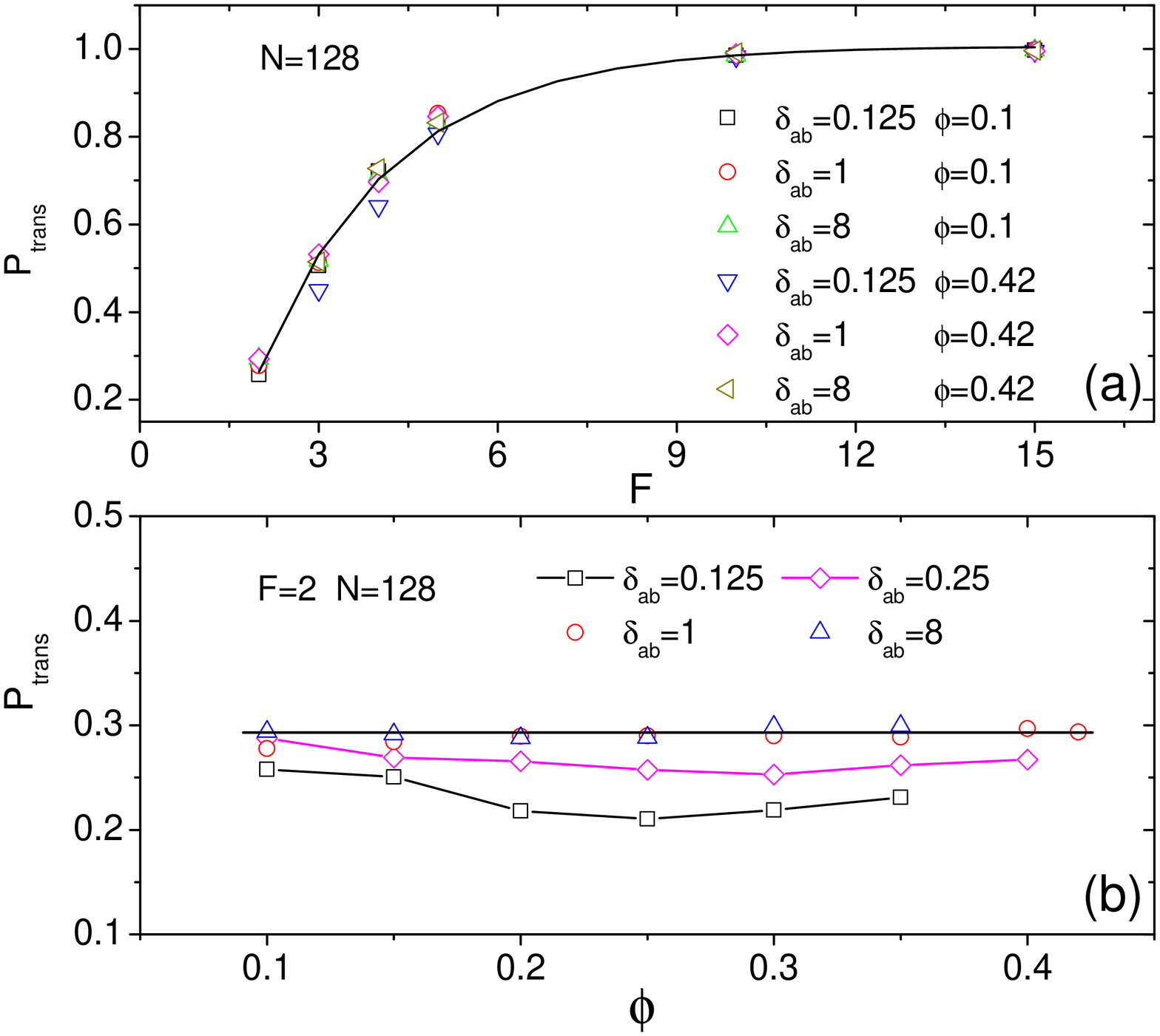}
\caption{(a) Translocation probability as function of the driving force $F$ for $N=128$ and different shapes of the
ellipsoid $\delta_{ab}$ and densities of the chain $\phi$. The solid line with $P_{trans}=-1.8125 e^{\frac{-F}{2.2416}}+1.0067$ is drawn to guide the eyes.
(b) Translocation probability as function of  $\phi$ for
$N=128$, $F=2$ and different $\delta_{ab}$.
        }
\label{Fig2}
\end{figure}

The translocation probability, $P_{trans}$, refers to the ratio of the successful translocation events to all tests at given parameters in
the simulation. As shown in Fig. \ref{Fig2}(a), with increasing the driving force $F$, $P_{trans}$ increases rapidly first, and then
slowly approaches saturation at larger driving force $F$. For strong driving forces $F\ge 10$, the effective driving forces $F(1-f(\phi)/F)$
are dominant, and the shape of the ellipsoid doesn't matter for $P_{trans}$. However, for weak driving forces $F\le 5$ the entropic resisting
forces $f(\phi)$ are important, and interestingly we observe that $P_{trans}$ depends on the shape of the ellipsoid.

To see the shape-dependence of $P_{trans}$ more clearly, we show the $P_{trans}$ as a function of $\phi$ under $F=2$ for different
$\delta_{ab}$, see Fig. \ref{Fig2}(b). $P_{trans}$ is almost constant for $ \delta_{ab} \ge 1$, independent of
$\phi$. However, $P_{trans}$ has weak $\phi$ dependence and decreases with decreasing $\delta_{ab}$ for $ \delta_{ab} < 1$. Obviously,
it is more difficult to translocation through major-axis direction of the ellipsoid, implying stronger entropic resisting force.
This is because at the early stage of the translocation process the translocated monomers are more confined for
$\delta_{ab} < 1$ compared with the cases for $\delta_{ab} \geq 1$.

\subsubsection{Effects of $\delta_{ab}$ and $\phi$ on the translocation time}

\begin{figure}
\includegraphics*[width=\figurewidth]{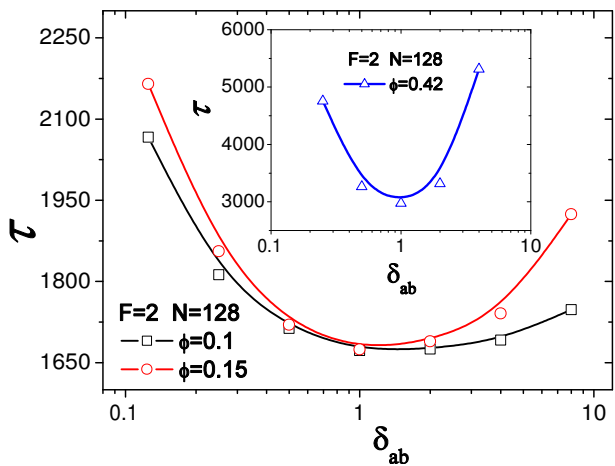}
\caption{Translocation time $\tau$ as a function of
the aspect ratio of the ellipsoid $\delta_{ab}$ for $N = 128$ and $F
= 2$ under the density of chain $\phi = 0.1$ and $0.15$. The inset
shows the case for the strong confinement $\phi = 0.42$.
        }
\label{Fig3}
\end{figure}

With increasing or decreasing $\delta_{ab}$ from $\delta_{ab}=1$ (sphere), the ellipsoid is more anisotropic.
Fig. \ref{Fig3} shows the translocation time $\tau$ as a function of $\delta_{ab}$ for different $\phi$. First, we find that with increasing
the anisotropy of the ellipsoid, $\tau$ always increases for different $\phi$.
This result indicates that compared with an isotropic confinement (sphere), an anisotropic confinement (ellipsoid) with the same volume
slows down the translocation velocity as observed by Ali. \emph{et al} \cite{Ali1}.

Figure \ref{Fig3} further demonstrates that it takes different time for polymer translocation into the same ellipsoid
through major-axis and minor-axis directions, depending on the average density of the whole chain in the ellipsoid, $\phi$.
For $\phi=0.1$ and 0.15, translocation into the same ellipsoid along the minor-axis direction  is faster.
However, for $\phi=0.42$ the results is opposite as shown in the inset of Fig. \ref{Fig3}.

\begin{figure}
\includegraphics*[width=\narrowfigurewidth]{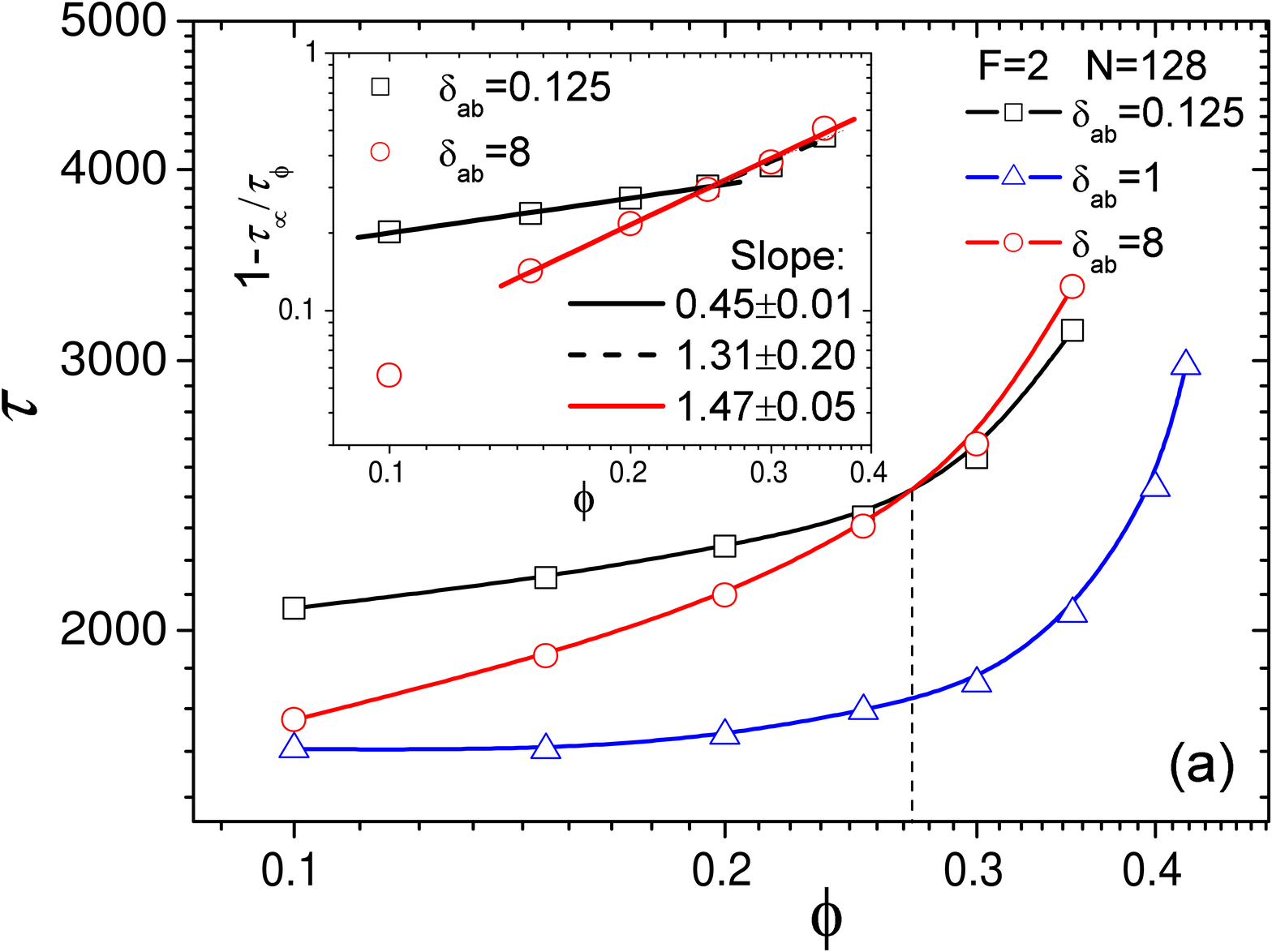}
\includegraphics*[width=\figurewidth]{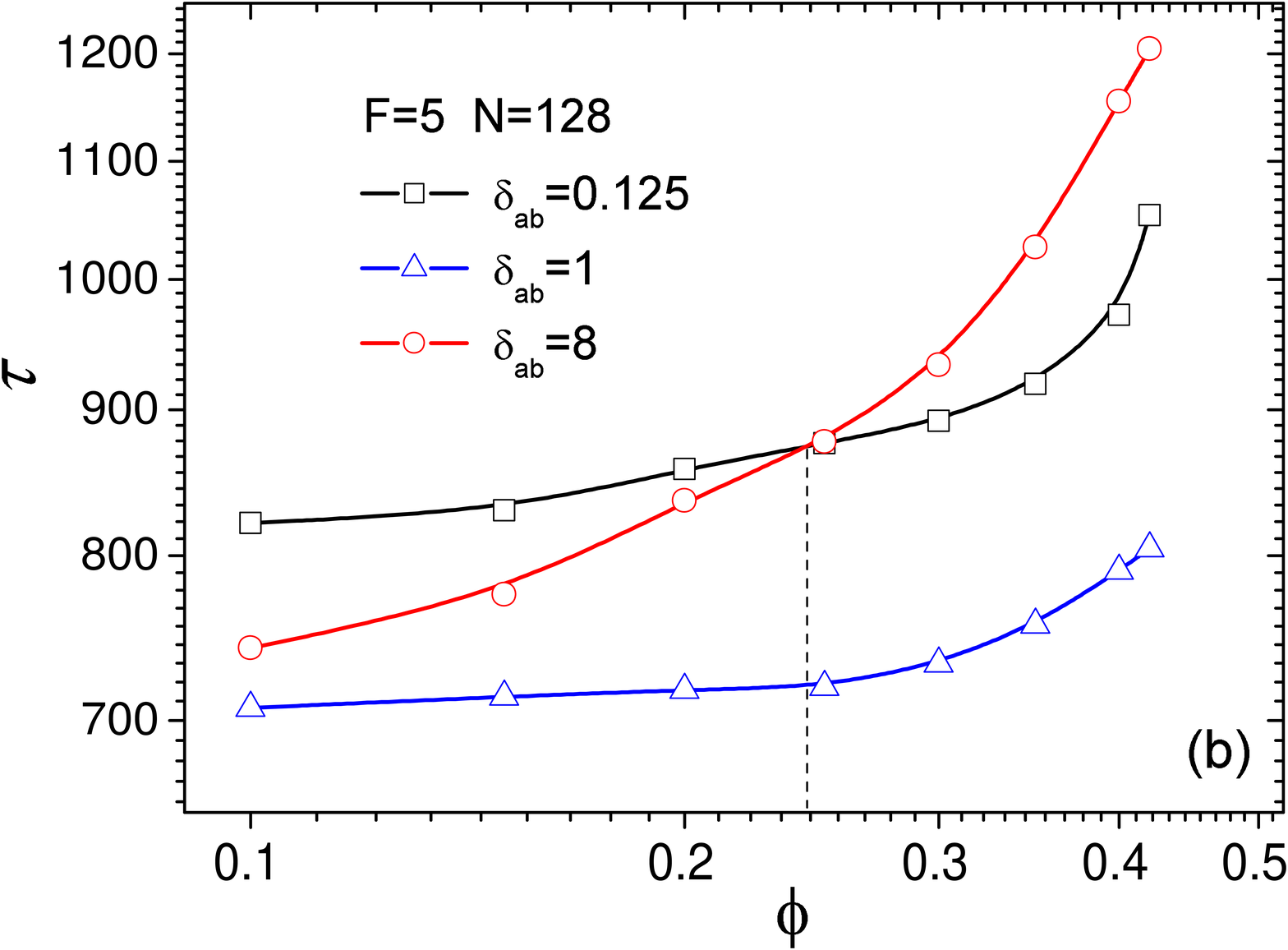}
\caption{Translocation time $\tau$ as a function of the density of the chain $\phi$ for $N =128$
and different aspect ratio of the ellipsoid $\delta_{ab}$ under: (a) the driving force $F = 2$ and (b) the the driving force
$F = 5$. The inset shows $1 - (\tau_{\infty}/\phi)\sim f(\phi)/F$ as a function of $\phi$.
        }
\label{Fig4}
\end{figure}

To give more details, we investigate the translocation time as a function of $\phi$ for $\delta_{ab}=0.125$, 1 and 8, respectively.
As shown in Fig. \ref{Fig4}, for all values of $\phi$, translocation into sphere is faster for both $F=2$ and 5.
For $\delta_{ab}=0.125$ and 1, $\tau$ initially increases slowly, then increases very rapidly with the increasing $\phi$. However, for
$\delta_{ab}=8$, $\tau$ initially increases a little faster compared with $\delta_{ab}=0.125$. As a main result, we find that there
exists a critical chain density, denoted as $\phi_c$. For $\phi<\phi_c$, translocation along the minor-axis direction is faster, while translocation
along the major-axis direction is faster for $\phi>\phi_c$, as shown in Fig. \ref{Fig4}(a) for $F=2$ and Fig. \ref{Fig4}(b) for $F=5$.
With increasing $F$, $\phi_c$ moves to smaller value, such as $\phi_{c} \approx 0.265$ for $F=2$ and 0.242 for $F=5$.

To understand the effect of the anisotropy of the ellipsoid on the translocation dynamics, we take into account the effective driving forces $F(1-f(\phi)/F)$.
Based on Eq. (\ref{eq9}), we calculate the average resisting force. As shown in the inset of Fig. \ref{Fig4}(a) for $F=2$, for $\phi<\phi_c$ the reduced
resisting force $f(\phi)/F$ for $\delta_{ab}=0.125$ is larger than that for $\delta_{ab}=8$. Thus, the effective driving forces $F(1-f(\phi)/F)$ for
$\delta_{ab}=8$ is stronger than that for $\delta_{ab}=0.125$, leading to a faster translocation for $\delta_{ab}=8$ as observed. However, for $\phi>\phi_c$
the effective driving forces for $\delta_{ab}=0.125$ is a little higher, resulting in faster translocation for $\delta_{ab}=0.125$.
For $\delta_{ab}=0.125$, we find that $f(\phi)\sim \phi^{0.45}$ for $\phi<\phi_c$ because the translocated monomers don't have enough time to diffuse away from
the entrance, leading to a high density of the translocated chain. For $\phi>\phi_c$ it crosses over to $f(\phi)\sim \phi^{1.31}$.
For $\delta_{ab}=8$, $f(\phi)\sim \phi^{1.47}$ is observed.
%

\begin{figure}
\includegraphics*[width=\narrowfigurewidth]{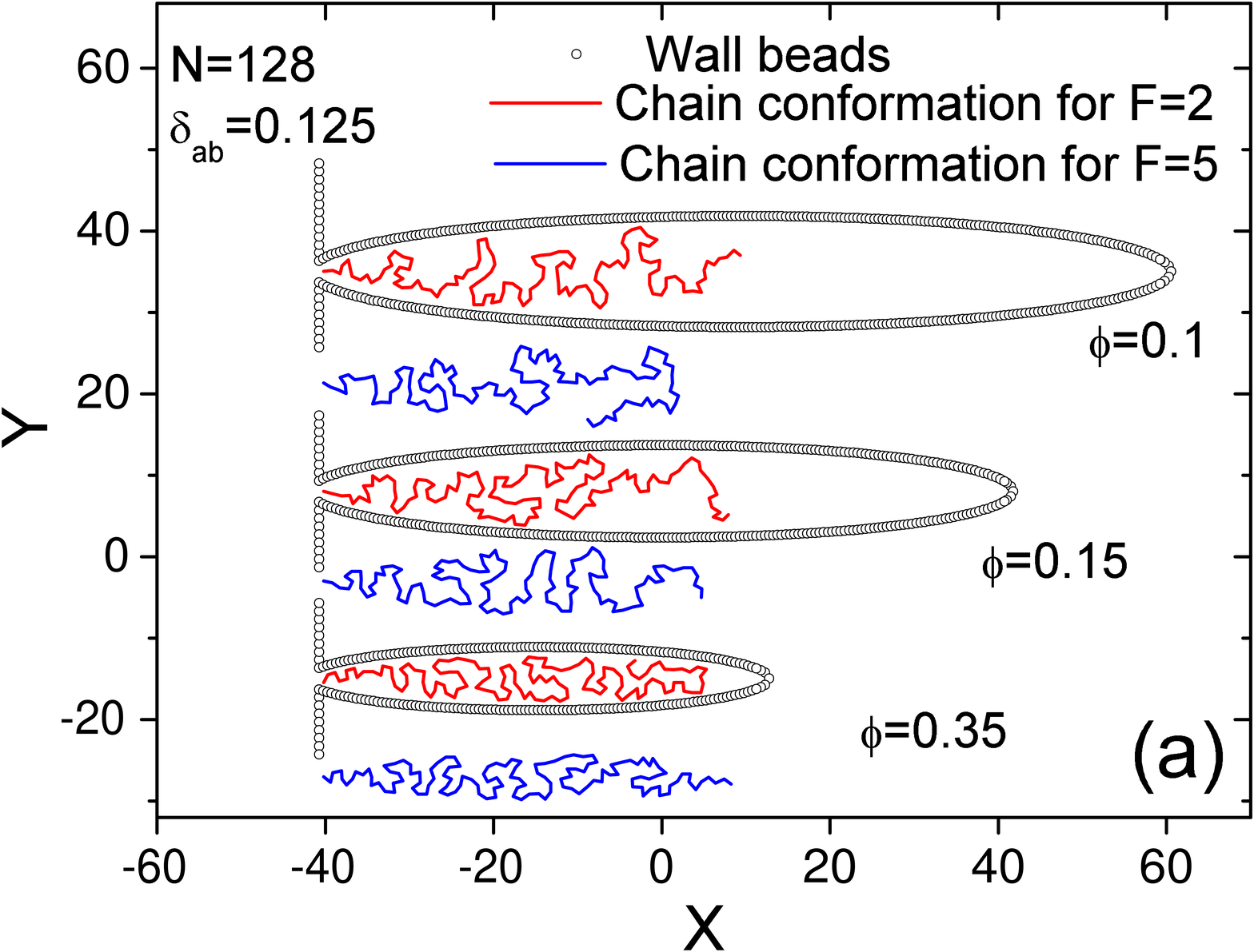}
\includegraphics*[width=\figurewidth]{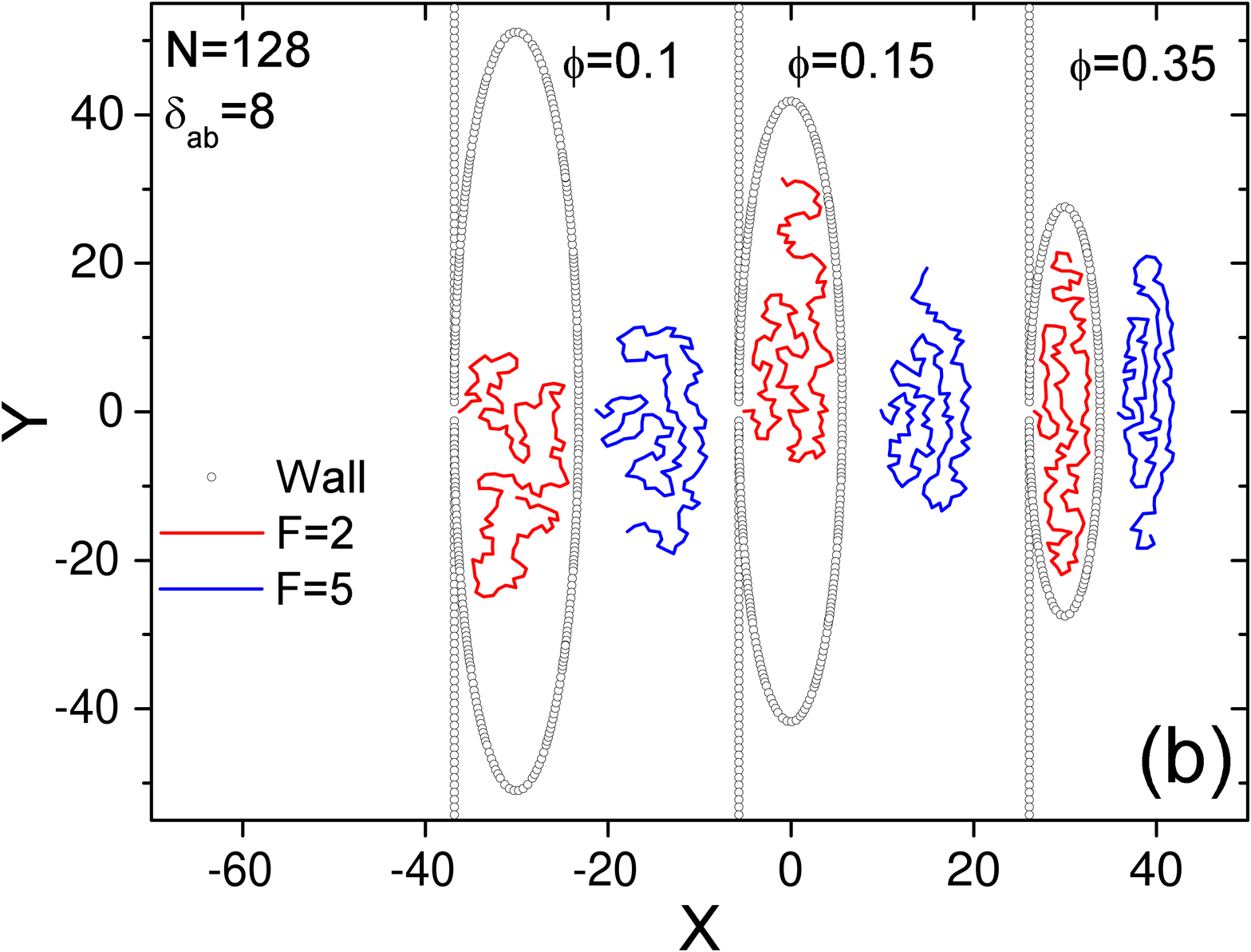}
\caption{(Color online) The chain conformation at the moment after the translocation for the chain length $N = 128$, and different
density of the chain $\phi$ under the driving force $F = 2$ and $5$: (a) the aspect ratio of the ellipsoid $\delta_{ab} = 0.125$ and (b)
$\delta_{ab} = 8$.
        }
\label{Fig5}
\end{figure}

In order to shed light on the variation of the average resisting force for $\delta_{ab}=0.125$ and 8, we give the chain conformation right after the translocation
for the chain $N = 128$ and different $F$.
As shown in Fig. \ref{Fig5}(a), for $\phi<\phi_c$ the translocated chain is more confined for $\delta_{ab}=0.125$ compared with $\delta_{ab}=8$, which
leads to a stronger resisting force as observed. However, for $\phi>\phi_c$ the translocated chain forms regular folded chain conformation for $\delta_{ab}=8$ as shown in Fig. \ref{Fig5}(b),
resulting in a stronger resisting force as we calculated. For larger driving forces, the translocated chain is easier to form folded chain conformation. For $\phi=0.35$, there even occur quadruple layers in the chain configuration for $F=5$, compared with triple layers for $F=2$. This leads to a smaller $\phi_c$ as shown in Fig. \ref{Fig4}.
In our previous study \cite{Luo10}, we also observed the obvious chain folding for polymer translocation into a fluidic channel through a nanopore under the strong driving force.
Compared with the translocation into a channel with parallel walls of the same size as the minor axis of the ellipsoid, the translocation into relatively thin ellipsoids is slower and is
easier to form the folded chain.

\subsubsection{Typical translocation process and waiting time distributions}

\begin{figure}
\includegraphics*[width=\figurewidth]{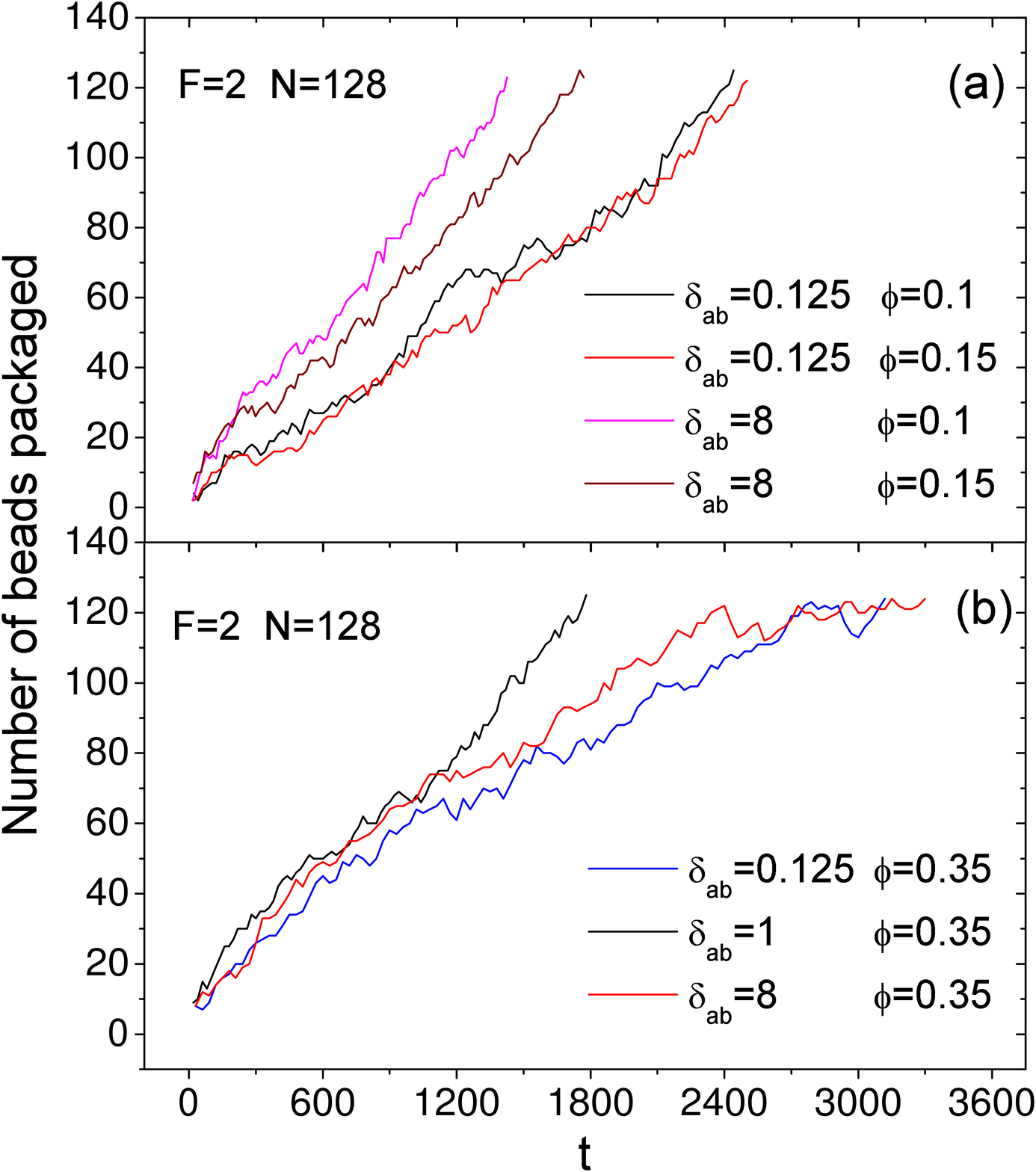}
\caption{Number of beads packaged against time for the chain length $N = 128$ and the driving force $F = 2$. (a)
$\phi = 0.1$ and 0.15 for different $\delta_{ab}$ and (b) $\phi = 0.35$ for different $\delta_{ab}$.
        }
\label{Fig6}
\end{figure}

To characterize the typical translocation process, in Fig. \ref{Fig6} we show the plot of the number of beads packaged against time for a successful
translocation event for $F=2$. For both $\delta_{ab} = 0.125$ and 8, translocation is faster for $\phi = 0.1$ than that for $\phi = 0.15$.
For higher $\phi$, such as $\phi=0.35$, the translocation velocity is much slower for $\delta_{ab} = 0.125$ and 8 than that for $\delta_{ab} = 1$
at the late stage of the translocation process.
Particularly, it is easier to form the folded chain for $\delta_{ab}=8$ compared with  $\delta_{ab}=1$, and it takes longer time for last several monomers
to get packaged.

\begin{figure}
\includegraphics*[width=\figurewidth]{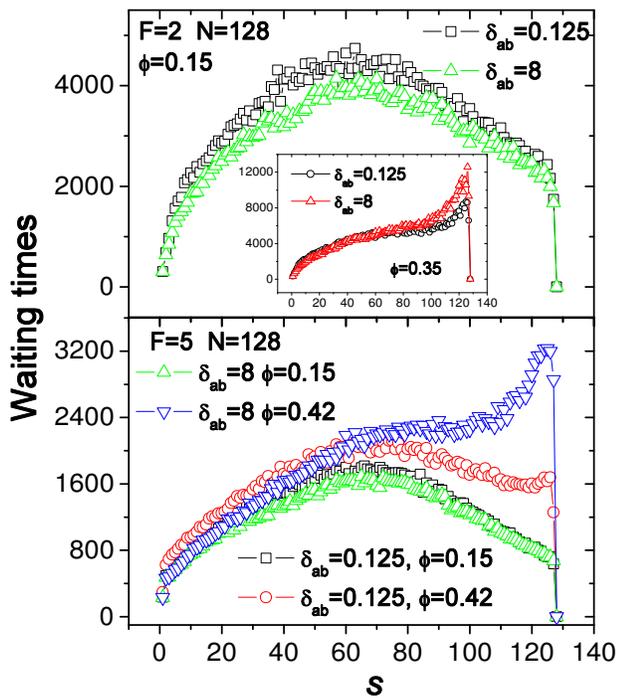}
\caption{Waiting time distribution for the chain
length $N =128$ and different $\phi$ and
$\delta_{ab}$. (a) $F = 2$ and (b)$F = 5$.
        }
\label{Fig7}
\end{figure}

To understand the translocation mechanisms, which is considerably affected by the nonequilibrium nature of the packaging process, it is
necessary to explore the dynamics of a single segment passing through the pore into the confinement. We numerically
calculate the waiting times for all monomers in a chain of length $N$. The waiting time of monomer $s$ is defined as the average time
between the events that monomer $s$ and monomer $s+1$ first pass through the exit of the pore.

Fig. \ref{Fig7} shows the waiting time distribution for the chain of length $N=128$ under different chain densities and different
$\delta_{ab}$ for $F=2$ and $5$.
For $\phi=0.15$, the waiting times for $s$ are longer for $\delta_{ab}=0.125$ than those
for $\delta_{ab}=8$ for $F=2$. This is because the reduced resisting force $f(\phi)/F)$ is stronger for $\delta_{ab}=0.125$, and it is important
for the effective driving force.
However, the reduced resisting force $f(\phi)/F)$ is almost negligible compared with the effective driving force
for $F=5$.  Thus, there is almost no difference in waiting times for $\delta_{ab}=0.125$ and 8. In addition, waiting time distributions are
almost symmetric with respect to the middle monomer $s=N/2$ as previously reported for translocation into an unconfined space \cite{Luo1,Luo2,Luo10}.

For $\phi=0.35$ and $F=2$, the waiting times increases rapidly after $s>N/2$ for both $\delta_{ab}=0.125$ and 8 due to the important role
of the resisting forces in translocation process.
For $\phi=0.42$ and $F=5$, the waiting times after $s>N/2$ are almost constant for $\delta_{ab}=0.125$. However, it still has a rapid increase for
$\delta_{ab}=8$, resulting from the formation of the folded chain conformation.

These detailed results further confirm the dependence of translocation dynamics on the shape of the ellipsoid and and the translocation direction.

\section{CONCLUSIONS} \label{chap-conclusions}

Using Langevin dynamics simulations in two dimensions, we investigate the dynamics of a flexible polymer translocation
into a confinement under a driving force $F$ through a nanopore. Particularly, we address the effects of shapes of a confinement on the translocation dynamics.
Compared with an isotropic confinement (sphere), an anisotropic confinement (ellipsoid) with the same volume slows down
the translocation velocity. With increasing the aspect ratio of the ellipsoid the translocation time increases.
We further find that it takes different time for polymer translocation into the same ellipsoid
through major-axis and minor-axis directions, depending on the average density of the whole chain in the ellipsoid, $\phi$.
For $\phi$ lower than a critical value $\phi_c$, the translocation through minor axis is faster, and vice versa.
These complicated behaviors are interpreted by the degree of the confinement and anisotropic confinement induced folding of the translocated chain.

In the future, it is very necessary to take into account the chain stiffness on the translocation dynamics. We expect that the anisotropic
confinement will play a more important role.

\begin{acknowledgments}
This work is supported by the National Natural Science Foundation of China (Grants No. 21225421, Grants No. 21174140 and No.
21074126) and the `` Hundred Talents Program '' of Chinese Academy of Science.
\end{acknowledgments}

\end{document}